\documentclass[aps, prx, reprint, twocolumn, notitlepage, showpacs, floatfix, superscriptaddress, longbibliography]{revtex4-1}

\usepackage{silence}
\usepackage{amsthm,amsmath,amsfonts,amssymb,verbatim, color}
\usepackage{graphicx}
\usepackage[percent]{overpic}
\usepackage{bm}
\usepackage{epsfig,slashed}
\usepackage{epstopdf}
\usepackage{lipsum}
\usepackage{float}
\usepackage{mathtools}
\usepackage{appendix}
\usepackage[colorlinks=true,citecolor=blue,linkcolor=magenta,urlcolor=blue]{hyperref}
\usepackage{natbib}
\usepackage{graphics} 
\graphicspath{ {figures/} }
\usepackage[caption=false]{subfig}
\usepackage[mathscr]{euscript}
\usepackage[export]{adjustbox}
\usepackage[bottom]{footmisc}
\usepackage{textcomp}
\usepackage{braket}
\makeatletter
\newsavebox\myboxA
\newsavebox\myboxB
\newlength\mylenA
\newcommand*\xoverline[2][0.75]{%
    \sbox{\myboxA}{$\m@th#2$}%
    \setbox\myboxB\null
    \ht\myboxB=\ht\myboxA%
    \dp\myboxB=\dp\myboxA%
    \wd\myboxB=#1\wd\myboxA
    \sbox\myboxB{$\m@th\overline{\copy\myboxB}$}
    \setlength\mylenA{\the\wd\myboxA}
    \addtolength\mylenA{-\the\wd\myboxB}%
    \ifdim\wd\myboxB<\wd\myboxA%
       \rlap{\hskip 0.5\mylenA\usebox\myboxB}{\usebox\myboxA}%
    \else
        \hskip -0.5\mylenA\rlap{\usebox\myboxA}{\hskip 0.5\mylenA\usebox\myboxB}%
    \fi}
\makeatother

\begin{document}

\newtheorem{theorem}{Theorem}
\newtheorem{property}{Property}
\newcommand{\tr}{\mathop{\mathrm{Tr}}}
\newcommand{\bsigma}{\boldsymbol{\sigma}}
\newcommand{\re}{\mathop{\mathrm{Re}}}
\newcommand{\im}{\mathop{\mathrm{Im}}}
\newcommand{\diag}{\mathrm{diag}}
\newcommand{\sign}{\mathrm{sign}}
\newcommand{\sgn}{\mathop{\mathrm{sgn}}}
\newcommand{\mb}{\bm}
\newcommand{\ua}{\uparrow}
\newcommand{\da}{\downarrow}
\newcommand{\ra}{\rightarrow}
\newcommand{\la}{\leftarrow}
\newcommand{\mc}{\mathcal}
\newcommand{\bs}{\boldsymbol}
\newcommand{\lra}{\leftrightarrow}
\newcommand{\nn}{\nonumber}
\newcommand{\half}{{\textstyle{\frac{1}{2}}}}
\newcommand{\mf}{\mathfrak}
\newcommand{\MF}{\text{MF}}
\newcommand{\IR}{\text{IR}}
\newcommand{\UV}{\text{UV}}

\renewcommand{\i}{\mathop{\mathrm{i}}}
\renewcommand{\b}[1]{{\boldsymbol{#1}}}

\def\bibsection{\section*{\refname}} 

\def\II{\hbox{$1\hskip -1.2pt\vrule depth 0pt height 1.6ex width 0.7pt\vrule depth 0pt height 0.3pt width 0.12em$}}
\DeclareGraphicsExtensions{.png}

\title{The tunable 0-\texorpdfstring{$\pi$}{Lg} qubit: Dynamics and Relaxation}

\author{Garima Rajpoot, Komal Kumari, Sandeep Joshi}
\affiliation{Theoretical Nuclear Physics and Quantum Computing Section, Nuclear Physics Division,\\ Bhabha Atomic Research Centre, Mumbai 400085, India}

\author{Sudhir R. Jain}
\affiliation{Theoretical Nuclear Physics and Quantum Computing Section, Nuclear Physics Division,\\ Bhabha Atomic Research Centre, Mumbai 400085, India}
\affiliation{Homi Bhabha National Institute, Anushakti Nagar, Mumbai 400094, India}
\affiliation{UM-DAE Centre for Excellence in Basic Sciences, University of Mumbai, Vidyanagari campus, Mumbai 400098, India}

\begin{abstract}
We present a systematic treatment of a 0-\texorpdfstring{$\pi$}{Lg} qubit in the presence of a time-dependent external flux. A gauge-invariant Lagrangian and the corresponding Hamiltonian is obtained. The effect of the flux noise on the qubit relaxation is obtained using perturbation theory. Under a  time-dependent drive of sinusoidal form, the survival probability, and transition probabilities have been studied for different strengths and frequencies. The driven qubit is shown to possess coherent oscillations among two distinct states for a weak to moderate strength close to resonant frequencies of the unperturbed qubit. The parameters can be chosen to prepare the system in its ground state. This feature paves the way to prolong the lifetime by combining ideas from weak measurement and quantum Zeno effect. We believe that this is an important variation of a topologically protected qubit which is tunable.       
\end{abstract}
\maketitle
\section{Introduction}
The scalability and reliability of the quantum computation demands development of ideas for protection of qubits against decoherence and noise \cite{nielsen2002quantum,kitaev2002classical,preskill2018quantum}. An important circuit design employing the superconducting Josephson junctions is the  current mirror qubit, and related $0-\pi$ qubit \cite{kitaev2006protected}. The circuit for the $0-\pi$ qubit is so-called due to the existence of two nearly degenerate ground states, localized at the superconducting phase difference between two leads near $\theta = 0$ and $\theta = \pi$ \cite{brooks2013protected}. Studies on coherence properties of the $0-\pi$ qubit show that it exhibits suppression of dephasing and relaxation in a significant way \cite{groszkowski2018coherence}. However, the regime of parameter values where it can be used to encode protected information is challenging to achieve currently. A practical implementation of this qubit has been realized recently \cite{gyenis2021experimental} where parameter regime is such that the ground state degeneracy is lifted but the protection from noise was still good. This interesting work has brought a novel possibility for a qubit with intrinsic protection for future designs. Employing an idea rooted deeply into nonlinear science, protection of qubits can be realized by choosing parameters respecting the condition of nonlinear resonances \cite{saini2020protection}. When a nonlinear system is perturbed, the invariant surfaces break, giving way to stable (elliptic) points and unstable (hyperbolic) points. The parameters can be so chosen that the system resides close to a stable point, protected by the separatrix passing through an unstable point. Various other ideas for intrinsic error protection have been well-studied and developed \cite{douccot2012physical,bell2014protected,gladchenko2009superconducting,smith2020superconducting,kalashnikov2020bifluxon}.     

The standard quantization procedure of a quantum circuit requires formulation of the  Hamiltonian description of the given circuit. In the presence of time-independent external flux this procedure leads to gauge dependent terms in the resulting Hamiltonian. Allowing for small fluctuations in the external flux may lead to ambiguous predictions for the physical observables. Thus external constraints are required to eliminate the gauge dependent terms from the Hamiltonian of the circuit itself. These constraints are termed as \textit{irrotational} constraints and they can be generalized for any single loop and multiloop quantum circuit. Imposing these irrotational constraints results in a Hamiltonian which is gauge invariant and thus gives rise to consistent predictions in the presence of time-dependent external flux  \cite{you2019circuit}.  

In this paper we carry out the procedure of circuit quantization of the $0-\pi$ qubit in the presence of a time-dependent external flux in a systematic manner. The $0-\pi$ qubit can be viewed as a quantum circuit  with three distinct loops, threaded by time-dependent external fluxes as shown in Fig. \ref{fig1}. A recent experimental realization of the $0-\pi$ qubit (see Fig. 3 of Ref.\cite{gyenis2021experimental})) gives a valid justification for the relevance of our toy model with three flux loops. To begin with, we obtain the gauge-invariant Hamiltonian of the circuit by enforcing irrotational constraint. The numerical diagonalization of the time-independent part of the Hamiltonian in the charge basis yields the eigenspectrum of the $0-\pi$ qubit. We then evaluate the qubit relaxation rate and corresponding transitions induced by the fluctuations in the external flux which allows us to set the optimal flux tuning for the qubit. Further, we present results arising due to driving the qubit, studying thereby the survival probability of an initially prepared state. The transition probabilities among the low-lying states are also studied. This allows us to conclude that the hierarchy of the states in the unperturbed $0-\pi$ qubit is preserved and that the definition of well-separated ``0" and ``1" states continues even in the presence of a drive. Hence, we have a protected qubit which is also tunable.  

\section{Gauge-invariant Lagrangian}

\begin{figure}[b!]
    \begin{center}
    \includegraphics[width=0.4\textwidth]{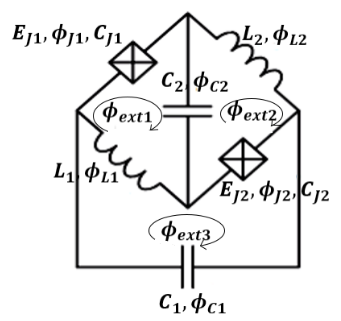}
    \caption{\small{Circuit for $0-\pi$ qubit with time dependent external flux, consisting of three loops. The external fluxes across the three loops are $\phi_{ext1}$, $\phi_{\rm ext2}$ and $\phi_{\rm ext3}$, respectively. $E_{\rm J1}$ and $E_{\rm J2}$ are the Josephson junction energies and the junction capacitances are $C_{\rm J1}$ and $C_{\rm J2}$, respectively. The flux passing through junction 1 is  $\Phi_{\rm J1}$ and that passing through junction 2 is $\Phi_{\rm J2}$.}}
    \label{fig1}
    \end{center}
\end{figure}

The $0-\pi$ qubit can be realized as a circuit consisting of two Josephson Junctions, two inductors and two capacitors arranged as shown in Fig. \ref{fig1}. We analyze the circuit of the $0-\pi$ qubit with the effect of time-dependent external flux. For the time-dependent case, to avoid inconsistencies, we employ the generalized method of circuit quantization, presented recently \cite{you2019circuit}. For the time-independent case, a considerable simplification occurs as one quantizes the circuit: one degree of freedom decouples \cite{dempster2014understanding} and there appears a simpler, lower-dimensional effective Lagrangian. Further, one may also employ Born-Oppenheimer approximation to reduce it to a one-dimensional problem with $\pi$-periodic potential \cite{shen2015theoretical}, eigenspectrum of which gives the qubit levels. However, once the system is explicitly time-dependent, the analysis has to be more generally carried out. 

We begin by the construction of the Lagrangian for the circuit. The kinetic energy terms in this Lagrangian correspond to capacitive energies. The potential energy is composed of the energies of all inductive elements, including those associated with Josephson junctions \cite{vool2017introduction}.

Let the branch fluxes corresponding to the Josephson junctions, capacitors and inductors be denoted by $\Phi_{\rm J1}, \Phi_{\rm J2}, \Phi_{\rm L1}, \Phi_{\rm L2}, \Phi_{\rm C1}, \Phi_{\rm C2}$ (Fig. \ref{fig1}). Let ${\Phi}_{\rm ext1}$, ${\Phi}_{\rm ext2}$ and ${\Phi}_{\rm ext3}$ denote the external fluxes threading the three loops in the circuit. We have six variables ($N$) corresponding to the branch fluxes and three constraints ($F$) corresponding to the loops. Hence, the circuit has three degrees ($(N-F)$) of freedom. We denote these by $\widetilde{\Phi}_i$, $i = 1, 2, 3$ such that their relation with branch flux vector may be expressed as \cite{you2019circuit}:
\begin{equation} \label{eq1}
 \widetilde{\bm{\Phi}}= \textbf{M}\bm{\Phi},
\end{equation}
where branch flux vector $\bm{\Phi}$ is given by
\begin{alignat}{1} \label{eq2}
 \bm{\Phi} = 
  (\Phi_{\rm J1}, 
  \Phi_{\rm J2}, 
  \Phi_{\rm L1}, 
  \Phi_{\rm L2}, 
  \Phi_{\rm C1}, 
  \Phi_{\rm C2})^{\rm T},
\end{alignat}
with T denoting the transpose. $\textbf{M}$ is an $(N-F)\times N$ i.e. ($3\times 6$) matrix. The constraints from the fluxoid quantization and the Faraday’s law can be expressed in a matrix form as:
\begin{equation} \label{eq3}
{\Phi}_{\rm ext} = \textbf{R} \bm{\Phi},
\end{equation}
where $\textbf{R}$ is an $F\times N$ ($3\times6$) mesh matrix whose elements are defined as follows:
\begin{equation} \nonumber
    \textbf{R} = \begin{cases}
    1 & \text {if $\Phi_i$ lies along orientation of $\Phi_{ext,i}$ }\\
   -1 & \text {if $\Phi_i$ is opposite to orientation of $\Phi_{ext,i}$ }\\
    0 & \text {if $\Phi_i$ is not present in the loop}\\
    \end{cases}.
\end{equation}
For the circuit, the matrix $\textbf{R}$ is given by: 
\begin{alignat}{1} \label{eq4}    
\textbf{R} & = \begin{pmatrix}
  1 & 0 & 1 & 0 & 0 & -1  \\
  0 & -1 & 0 & -1 & 0 & 1\\
  0 & 1 & -1 & 0 & 1 & 0\\
  \end{pmatrix}.
\end{alignat}
We can write \eqref{eq1} and \eqref{eq3} in an augmented form by introducing:
\begin{equation}\label{eq5}
    \widetilde{\bm{\Phi}}_{+}= \begin{pmatrix}
\widetilde{\bm{\Phi}} \\
\bm{\Phi}_{\rm ext}
\end{pmatrix}, \quad
\textbf{M}_+ = \begin{pmatrix}
   \textbf{M} \\
   \textbf{R} 
\end{pmatrix},
\end{equation}
so that the combined equation becomes
\begin{equation}\label{eq6}
\widetilde{\bm{\Phi}}_+=\textbf{M}_+ \bm{\Phi},
\end{equation}
where the determinant $\lvert{\textbf{M}}_+\rvert\neq0$ . To identify the irrotational degrees of freedom, we turn to the kinetic energy
\begin{alignat}{1} \label{eq7}
\mathcal{L}_{\rm k} = \frac{1}{2}\dot{\bm{\Phi}}^T \textbf{C} \dot{\bm{\Phi}} 
   = \frac{1}{2}\dot{\widetilde{\bm{\Phi}}}_+^{\rm T} \textbf{C}_{\rm eff}\dot{\widetilde{\bm{\Phi}}}_+,
\end{alignat}
with
\begin{equation}\label{eq8}
 \textbf{C}_{\rm eff} = ({\textbf{M}_+}^{-1})^{\rm T} \textbf{C} {\textbf{M}_+}^{-1}.
\end{equation} 
We consider the symmetric case in which the two Josephson junctions, the two capacitors as well as the two inductors are identical to each other. We thus have, $E_{\rm J1}$= $E_{\rm J2}$=$E_{\rm J}$, $C_{\rm J1}$= $C_{\rm J2}$=$C_{\rm J}$, $C_{\rm C1}$=$C_{\rm C2}$=$C_{\rm C}$ and $L_1= L_2= L$.
For this case, the diagonal matrix $\textbf{C}$ is given by
 \begin{alignat}{1} \label{eq9}
 \textbf{C} &= {\rm Diagonal~}(C_{\rm J}, C_{\rm J}, C_{\rm L}, C_{\rm L}, C_{\rm C}, C_{\rm C}),
\end{alignat}
where $C_{\rm L}$ represents the auxiliary parallel capacitance associated with the inductors.  

The irrotational degrees of freedom are obtained by demanding
that the terms in the Lagrangian proportional to $\dot\Phi_{\rm ext}$ vanish, leading to the condition \cite{you2019circuit}
\begin{equation}\label{eq10}
    \textbf{R}\textbf{ C}^{-1}\textbf{M}^{\rm T} =0.
\end{equation}
The above condition, in the limit $C_{\rm L}\rightarrow 0$ results in the matrix $C_{\rm eff}$ to assume a block-diagonal form. As shown in \ref{appex-1}, this allows us to find the matrices  $\textbf{M}$, $\textbf{ M}_+$ and $\textbf{C}_{\rm eff}$. The resulting gauge invariant kinetic energy term in \eqref{eq7} is given by
\begin{alignat}{1}\label{eq11}
    \mathcal{L}_{\rm k}&(\Vec{\widetilde{\bm{\Phi}}},\dot{\Vec{\widetilde{\bm{\Phi}}}},t)=\frac{1}{4C_{\Sigma}}\Big[4C_{\Sigma} \big(C_\Sigma (\dot{\widetilde{\Phi}}_1^2+2\dot{\widetilde{\Phi}}_1\dot{\widetilde{\Phi}}_2+2\dot{\widetilde{\Phi}}_2^2)\nonumber \\
    &-2C_{\rm J}\dot{\widetilde{\Phi}}_2\dot{\widetilde{\Phi}}_3+  C_{\rm J}\dot{\widetilde{\Phi}}_3^2\big)+C_{\rm C} C_{\rm J}(\dot{{\Phi}}_{\rm ext1}(t)+\dot{{\Phi}}_{\rm ext3}(t))^2\Big],
\end{alignat}
where $\Vec{\widetilde{\bm{\Phi}}}$ and $\dot{\Vec{\widetilde{\bm{\Phi}}}}$ include all three components of $\widetilde{\bm{\Phi}}$ and $\dot{\widetilde{\bm{\Phi}}}$ and $C_\Sigma= C_{\rm C}+ C_{\rm J}$.

\section{Hamiltonian and Eigenspectrum}\label{sec-3}

The potential energy of the circuit includes the energy of the Josephson junctions and the energy stored in the inductors. This is given by 
\begin{alignat}{1}\label{eq12}
   \mathcal{H}_{\rm U}(\vec{\phi},t)&=-E_{\rm J}\cos{\phi_{\rm J1}}-E_{\rm J}\cos{\phi_{\rm J2}}+\frac{1}{2}E_{\rm L}\phi_{\rm L1}^2+\frac{1}{2}E_{\rm L}\phi_{\rm L2}^2,
\end{alignat}
where $E_{\rm L} = (\Phi_0/2\pi)^2/L$. The phase variables $\phi$ are related to the flux variables $\Phi$ as: $\phi=2\pi\Phi/\Phi_0$ where, $\Phi_0 = h/2e$ is the flux quantum, $h$ is the Planck's constant and $e$ is electron charge. The branch fluxes can be written in terms of degrees of freedom of the circuit and the external fluxes using  \eqref{eq6}, where the matrix $\textbf{M}_{+}^{-1}$ is given by \eqref{eqa7}. Thus, the potential energy of the circuit becomes
\begin{align}\label{eq13}
    \mathcal{H}_{\rm U}(\vec{\widetilde{\phi}},t)&=-2E_{\rm J}\cos(\widetilde{\phi}_2-\widetilde{\phi}_3)\cos(\widetilde{\phi}_1+\widetilde{\phi}_2+\xi(\phi_{\rm e1}(t)\nonumber\\
    &+\phi_{\rm e3}(t))+\frac{E_{\rm L}}{2}\left[\frac{1}{2}(\phi_{\rm e1}(t)-\phi_{\rm e3}(t))-\widetilde{\phi}_3\right]^2 \nonumber\\
    &+\frac{E_{\rm L}}{2}\left[2\widetilde{\phi}_2-\widetilde{\phi}_3+\frac{1}{2}(\phi_{\rm e1}(t)+2\phi_{\rm e2}(t)+\phi_{\rm e3}(t))\right]^2,
\end{align}
where $\xi = C_{\rm C}/2C_{\Sigma}$ and $\vec{\widetilde{\phi}}$ includes the three dynamic ${\widetilde{\phi}}$ variables. Here, $\phi_{\rm e,i}$(t)$ = \phi_{\rm ext, i}$(t) denote the phase variables corresponding to the external fluxes $\Phi_{\rm ext,i}$(t). The above procedure allows us to construct a gauge-invariant Lagrangian in terms of canonically conjugate variables. In the next step we proceed to obtain the circuit Hamiltonian by performing the Legendre transformation. The charge variables $q_i$ canonically conjugate to the flux variables $\widetilde{\Phi}_i$ are given by
\begin{alignat}{1}\label{eq14}
    q_1 &= 2C_{\Sigma}(\dot{\widetilde{\Phi}}_1+\dot{\widetilde{\Phi}}_2), \nonumber\\
    q_2 &=2C_{\Sigma}(\dot{\widetilde{\Phi}}_1+2\dot{\widetilde{\Phi}}_2)-2C_{\rm J}\dot{\widetilde{\Phi}}_3, \nonumber\\
    q_3 &=-2C_{\rm J}(\dot{\widetilde{\Phi}}_2-\dot{\widetilde{\Phi}}_3).
\end{alignat}
To write the Hamiltonian, we have to express  $\dot{\widetilde{\Phi}}_i$ in terms of $q_i \equiv 2 e n_i$, where $n_i$ represents the number of cooper pairs. Using the \eqref{eq14}, we obtain 
\begin{alignat}{1}\label{eq15}
    \dot{\widetilde{\Phi}}_1 &= \frac{(C_{\rm C}+C_\Sigma)q_1-C_\Sigma (q_2+q_3)}{2C_{\rm C}C_\Sigma}, \nonumber\\
    \dot{\widetilde{\Phi}}_2 &= \frac{1}{2C_{\rm C}}(-q_1+q_2+q_3), \nonumber\\
    \dot{\widetilde{\Phi}}_3 &=\frac{1}{2C_{\rm C}C_{\rm J}}[C_{\rm J}(-q_1+q_2)+C_{\Sigma}q_3].
\end{alignat}
The Legendre transformation of \eqref{eq11} thus gives the Hamiltonian of the system:
\begin{equation}\label{eq16}
         H(\Vec{\widetilde\Phi},\vec{q},t) =\mathcal{H}_k(\vec{q},t)+\mathcal{H}_U(\Vec{\widetilde\Phi},t),
\end{equation}
where $\mathcal{H}_U(\Vec{\widetilde\Phi},t)$ is given by Eq.\eqref{eq13} and $\mathcal{H}_k(\vec{q},t)$, the Kinetic part of the Hamiltonian is:
\begin{alignat}{1}\label{eq17}
\mathcal{H}_{\rm k}(\vec{q},t) &= \frac{E_{\rm C \Sigma}}{4 C_{\rm C} C_{\rm J} C_\Sigma^2}\big( A n_1^2+Bn_2^2+Cn_3^2 \nonumber \\
& +Dn_1n_2+En_2n_3+F n_3n_1\big)+ G(t)\nonumber\\
&= {\mathcal H}_{\rm k0}(\vec{n}) +G(t),
\end{alignat}
where $E_{\rm C \Sigma} = e^2/2 C_{\Sigma}$, and the time-dependent term is given by
\begin{equation} \label{eq18}
    G(t)=\frac{1}{4 C_\Sigma}C_{\rm C} C_{\rm J}(\dot{{\phi}}_{\rm ext1}(t)+\dot{{\phi}}_{\rm ext3}(t))^2.
\end{equation}
The explicit forms of  $A$, $B$, $\dots$, $F$ are given in  \eqref{eqb1}.
It is evident that $\mathcal{H}_{\rm k}$ is quadratic in charge variables while $\mathcal{H}_{\rm U}$ is a nonlinear function in $\widetilde\phi_i$ and $\phi_{\rm ext,i}$. 

In the standard case of time-independent external flux, the Hamiltonian of the $0-\pi$ qubit has the form \cite{dempster2014understanding}:
\begin{align} \label{eq19}
    H(\theta, \phi)  =& -2 E_{\rm CJ} \partial_\phi^2 - 2 E_{\rm C \Sigma} \partial_\theta^2 \nonumber \\& -2E_{\rm J} \cos \theta \cos (\phi - \phi_{\rm ext}/2) + E_{\rm L}\phi^2.
\end{align}
It can be seen in this case that the potential term has two ridges at $\theta= 0$ and $\theta= \pi$ such  that the minima in the two ridges are staggered with respect to each other. For a specific set of parameters the lowest two energy levels of such a potential are nearly degenerate and the corresponding wave functions are localised along $\theta=0$ and $\theta= \pi$ ridges. In particular, we require the condition $E_{\rm L}, E_{\rm C \Sigma} \ll E_{\rm J}, E_{\rm CJ}$ for robust protection of the qubit against charge dispersion and disorder in the circuit parameters \cite{dempster2014understanding}. 

For our case, we choose three sets of parameters given by \\
Set 1: $E_{\rm C} = 1.2$, $E_{\rm CJ} = 4$, $E_{J} = 6$, $E_L= 0.038$, \\
Set 2: $E_{\rm C} = 0.15$, $E_{\rm CJ} = 10$, $E_{J} = 5$, $E_L= 0.13$, \\
Set 3: $E_{\rm C} = 0.65$, $E_{\rm CJ} = 1.75$, $E_{J} = 10.8$, $E_L= 0.79$.

\begin{table}[tbph!]
    \centering
    \begin{tabular}{|c|c|c|c|}
         \hline
           & Set 1 & Set 2 & Set 3 \\ [0.5ex] 
         \hline
        A & 8.329 & 1.190 & 4.390 \\
        B & 4.661 & 0.600 & 2.512 \\
        C &  17.969 & 35.488 & 8.281\\
        D & -9.323 & -1.198 & -5.024 \\
        E & 8.400 & 1.050 & 4.550 \\
        F & -8.506 & -1.051 & -4.614 \\
        \hline
    \end{tabular}
    \caption{Coefficients of the kinetic energy in Eq. \eqref{eq17} for the three sets of parameters are shown.}
    \label{tab1}
\end{table}

\begin{figure*}[t!]
    \centering
    \subfloat[]{\includegraphics[width=0.33\textwidth]{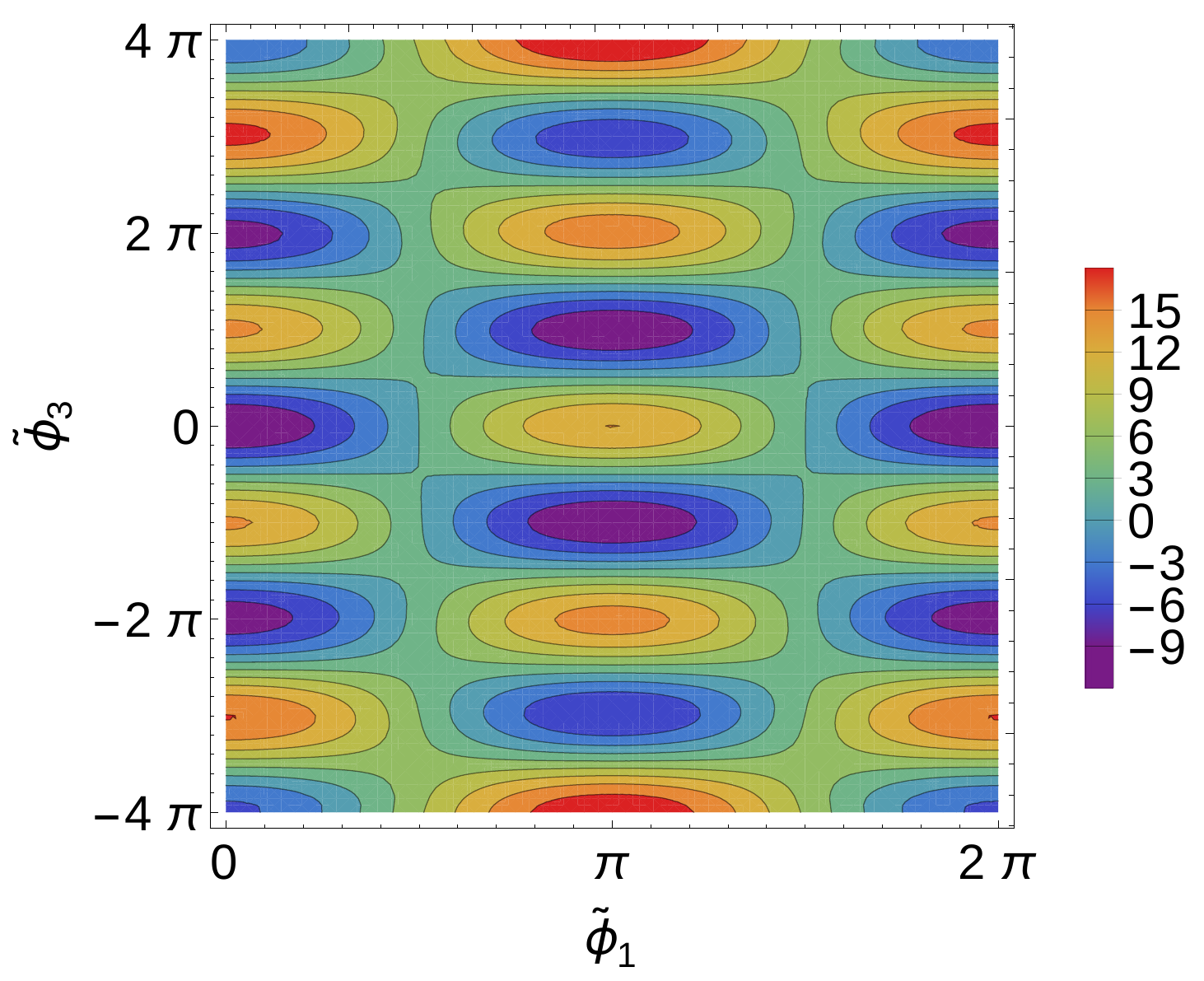}} 
    \subfloat[]{\includegraphics[width=0.33\textwidth]{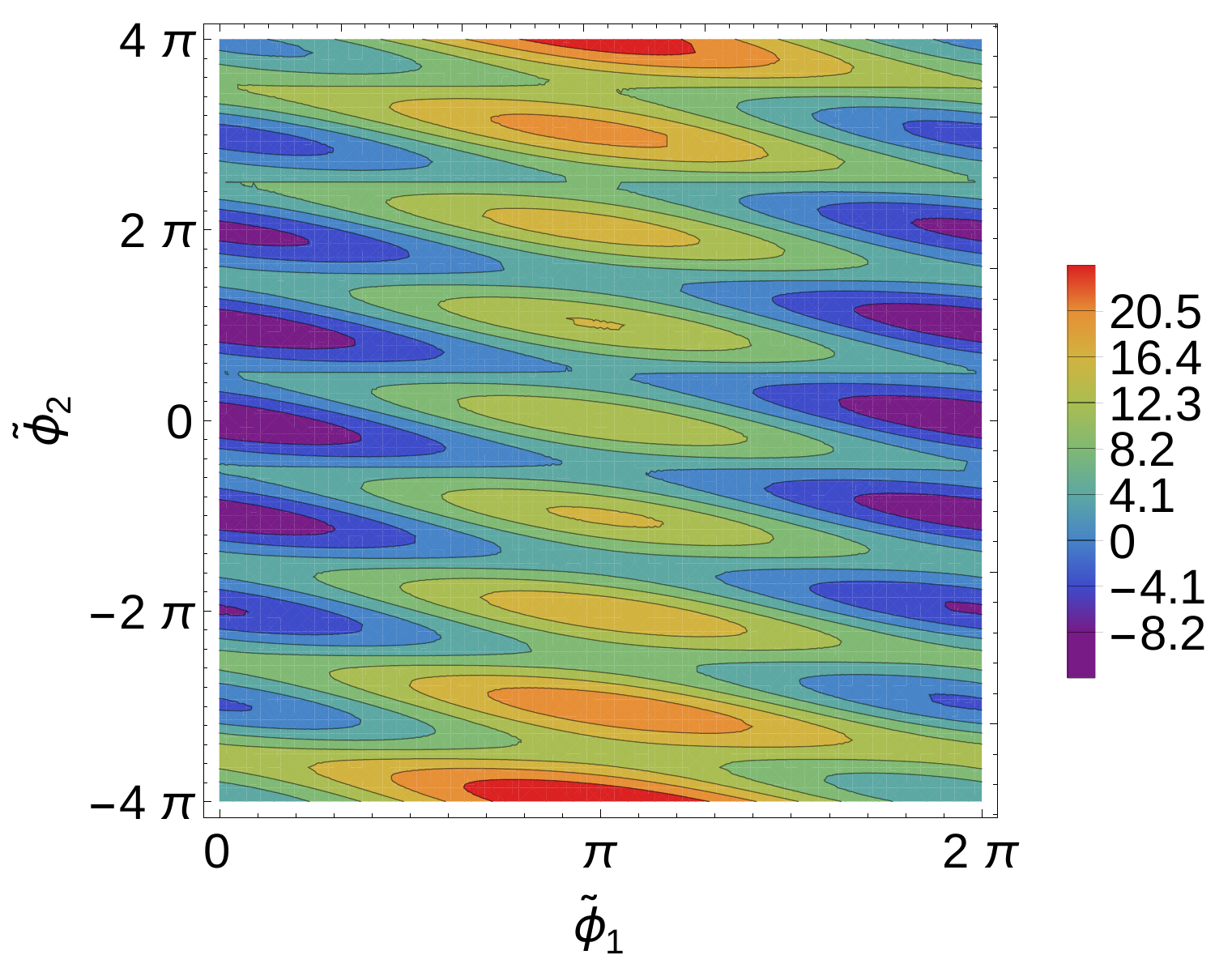}} 
    \subfloat[]{\includegraphics[width=0.33\textwidth]{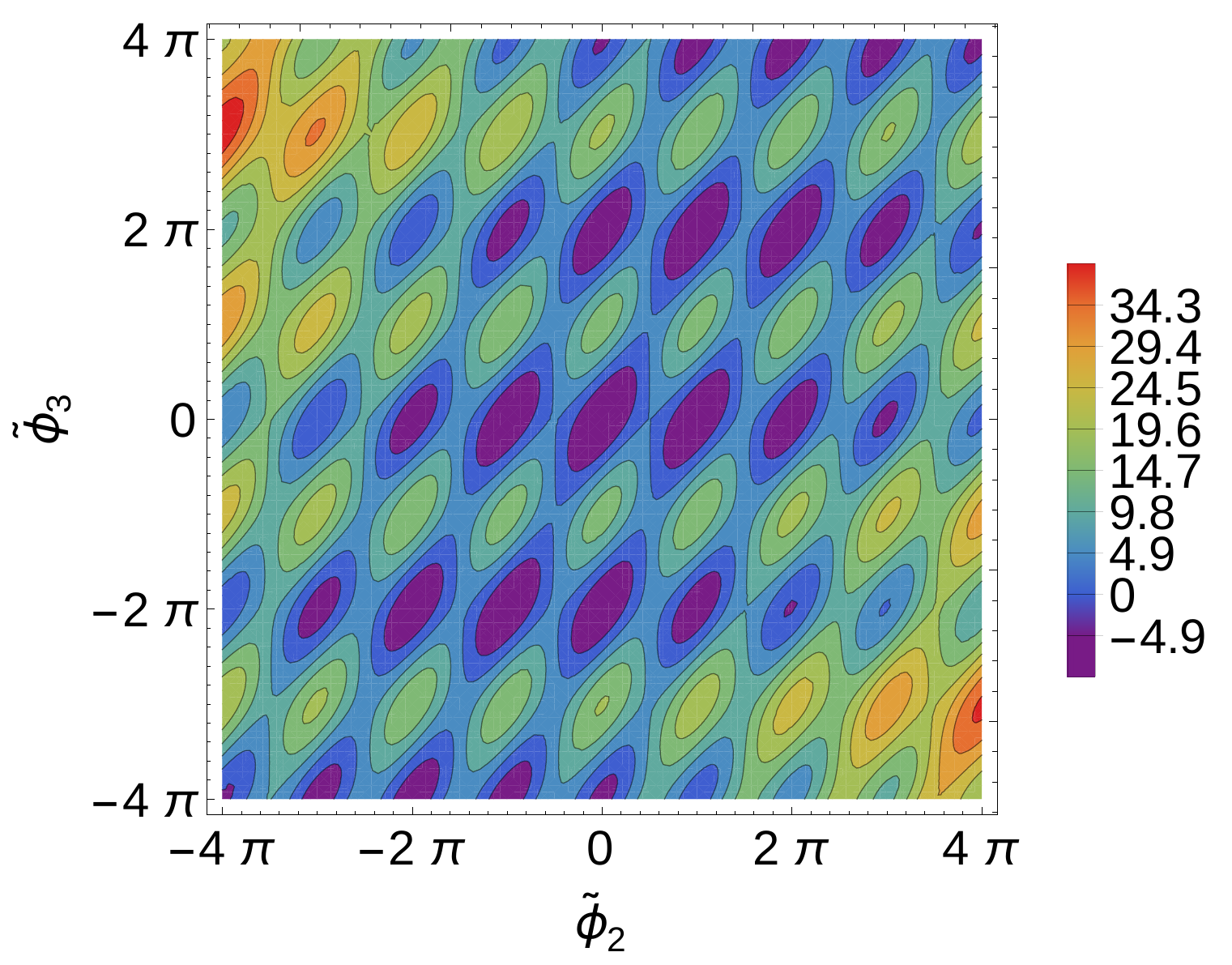}}
    \caption{Potential energy ( Eq. \eqref{eq13}) landscape for circuit parameters Set 1. The potential is periodic in the $\widetilde{\phi}_1$ variable and shows staggered minima in the $\widetilde{\phi}_1-\widetilde{\phi}_3$ plane (Fig. (a)). In the $\widetilde{\phi}_1-\widetilde{\phi}_2$ plane (Fig. (b)), since B $<$ A (Table \ref{tab1}) the wavefunction localization is more along $\widetilde{\phi}_2$ direction as compared to $\widetilde{\phi}_1$ direction. In Fig. (c), the potential is shown in the $\widetilde{\phi}_2-\widetilde{\phi}_3$ plane where again B $<$ C, which leads to delocalization of wavefunction along $\widetilde{\phi}_3$ direction. }
\label{fig2}
\end{figure*}

For the above sets of parameters we calculate the coefficients A, B, ..,F of the kinetic energy term \eqref{eq17} as shown in Table \ref{tab1}. These coefficients represent the effective mass tensor in the space of variables $(\widetilde{\phi}_1, \widetilde{\phi}_2, \widetilde{\phi}_3)$ and their magnitude is related to tunneling of the wavefunction along these directions. For instance, from Table \ref{tab1} we can see that for all of the sets $C \gg A, B$, which implies effective mass along $\widetilde{\phi}_3$ direction is much smaller compared to those along $\widetilde{\phi}_1$ and $\widetilde{\phi}_2$ direction. This leads to relatively larger delocalization of wavefunction along the $\widetilde{\phi}_3$ as compared to $\widetilde{\phi}_1, \widetilde{\phi}_2$ directions. 

In Fig. \ref{fig2}, we plot the potential \eqref{eq13} in the three different planes for the parameter Set 1. As can be seen from \eqref{eq13}, the potential is periodic in $\widetilde{\phi}_1$ and shows staggered minima in the $\widetilde{\phi}_1=0$ and $\widetilde{\phi}_1= \pi$ ridges. Also as discussed in the preceding paragraph the wavefunction is delocalized over multiple minima along $\widetilde{\phi}_3$ direction. Thus the $\widetilde{\phi}_1-\widetilde{\phi}_3$ plane is analogous to the standard $\theta-\phi$ plane of Ref. \cite{dempster2014understanding}. However, in our case due to the presence of an extra degree of freedom the potential assumes more complex form and the effective mass tensor (a 3 $\times $ 3 matrix with terms given by the coefficients, $A$ to $F$) determines the localization of the wavefunction along these three degrees of freedom.

Let us consider that for each  loop $i=1,2,3$, the external flux variable can be written as a sum of time-independent and time-dependent terms as,
\begin{equation}\label{eq20}
    {\phi}_{\rm ext, i}(t)=\phi_{\rm e,i}^{0}+\delta\phi_{\rm e, i}(t).
\end{equation}
where $\phi_{\rm e,i}^{0}$ is some fixed controllable value while $\delta\phi_{\rm e, i}(t)$ represents the time dependent fluctuations in the external flux. This allows us to re-write the Hamiltonian as $H(\Vec{\widetilde\Phi},\vec{q},t)=\mathcal{H}_0(\Vec{\widetilde\Phi},\vec{q})+\mathcal{H}(\Vec{\widetilde\Phi},\vec{q},t)$ where $\mathcal{H}_0(\Vec{\widetilde\Phi},\vec{q})$ denotes the time-independent part, and, $\mathcal{H}(\Vec{\widetilde\Phi},\vec{q},t)$ denotes the perturbed part. We imagine that there is no  perturbation before the time, $t = 0$, to facilitate the splitting of the Hamiltonian, in the same vein as in any standard treatment of linear response theory. 

In the Sec. \ref{sec:noise}, we shall employ these results to calculate the relaxation rate of qubit. Finally, in Sec. \ref{sec:survival}, we consider how an explicitly time-dependent sinusoidal drive affects the transition probabilities. 

Returning to the present discussion, a Taylor expansion of the total Hamiltonian about $t = 0$ (upto second order in t) allows us to realize this expression:
\begin{alignat}{1}\label{eq21}
    \mathcal{H}_0(\Vec{\widetilde\Phi},\vec{q})& = \mathcal{H}_{\rm k0}-2E_{\rm J}\cos(\widetilde{\phi}_2-\widetilde{\phi}_3)\bigg[\cos(\widetilde{\phi}_1+\widetilde{\phi}_2+\xi{\phi_{\rm e13}^{+}})\bigg]\nonumber\\
    &+\frac{E_{\rm L}}{2}\bigg[4\widetilde{\phi}_2^2+2\widetilde{\phi}_3^2-4\widetilde{\phi}_2\widetilde{\phi}_3+\frac{1}{4}{(\phi_{\rm e13}^{-})}^2-\widetilde{\phi}_3{\phi_{\rm e13}^{-}}\nonumber\\
    &+\frac{1}{4}({\phi_{\rm e123}^{+}})^2+(2\widetilde{\phi}_2-\widetilde{\phi}_3)({\phi_{\rm e123}^{+}})\bigg],
\end{alignat}
and
\begin{alignat}{1} \label{eq22}
    \mathcal{H}&(\Vec{\widetilde\Phi},\vec{q},t) = G(t)+2E_{\rm J}\cos(\widetilde{\phi}_2-\widetilde{\phi}_3)\bigg[\cos(\widetilde{\phi}_1+\widetilde{\phi}_2)\nonumber\\
    &\bigg\{\xi(\delta\phi_{\rm e13}^{+}(t))\sin(\xi\phi_{\rm e13}^{+})+\frac{\xi^2}{2}(\delta\phi_{\rm e13}^{+}(t))^2\cos(\xi\phi_{\rm e13}^{+})\bigg\}\nonumber\\
    &+\sin(\widetilde{\phi}_1+\widetilde{\phi}_2)\bigg\{\xi\delta\phi_{\rm e13}^{+}(t)\cos(\xi\phi_{\rm e13}^{+}) -\frac{\xi^2}{2}(\delta\phi_{\rm e13}^{+}(t))^2\nonumber\\
    &\sin(\xi\phi_{\rm e13}^{+})\bigg\}\bigg]+\frac{E_{\rm L}}{2}\bigg[\frac{1}{2}(\delta\phi_{\rm e13}^{-}(t))\phi_{\rm e13}^{-}-(\delta\phi_{\rm e13}^{-}(t))\widetilde{\phi}_3\nonumber\\
    &+\frac{1}{4}(\delta\phi_{\rm e13}^{-}(t))^2+\frac{1}{2}(\delta\phi_{\rm e123}^{+}(t))(\phi_{\rm e123}^{+})+(\delta\phi_{\rm e123}^{+}(t))\nonumber\\
    &(2\widetilde{\phi}_2-\widetilde{\phi}_3)+\frac{1}{4}(\delta\phi_{\rm e123}^{+}(t))^2\bigg].
\end{alignat}
where, ${\phi_{\rm e13}}^{(\pm)}=\phi_{\rm e1}^{0}\pm\phi_{\rm e3}^{0}$, ${\phi_{\rm e123}}^{(+)}=\phi_{\rm e1}^{0}+2\phi_{\rm e2}^{0}+\phi_{\rm e3}^{0}$, $\delta\phi_{\rm e13}^{\pm}(t) = \delta\phi_{\rm e1}(t)\pm\delta\phi_{\rm e3}(t)$ and $\delta\phi_{\rm e123}^{+}(t) = \delta\phi_{\rm e1}(t)+2\delta\phi_{\rm e2}(t)+\delta\phi_{\rm e3}(t)$.
To find the eigenspectrum, we have considered the adiabatic evolution of $\phi_{\rm ext}(t)$ with time. For the adiabatic evolution with time : $\phi_{\rm e,i}(t) = \phi_{\rm e,i}^0(1-e^{-\epsilon t})$. For small value of $\epsilon$ the evolution in $\phi_{\rm e,i}(t)$ is very small and linear with time which we have considered here to plot the eigenvalues with $\phi_{\rm ext,i}$.

We numerically diagonalize the Hamiltonian given by \eqref{eq21} in the charge basis. We define charge momentum operator $(\hat{n}_1,\hat{n}_2, \hat{n}_3)$ conjugate to the corresponding flux operator $(\hat{\widetilde{\phi}}_1,\hat{\widetilde{\phi}}_2,\hat{\widetilde{\phi}}_3)$ such that
\begin{equation}\label{eq23}
    \hat{n}_i = \frac{i}{2 \sqrt{\xi}} \big(\hat{a}_i^\dagger - \hat{a}_i\big), ~ i= 1,2,3,
\end{equation}
where $\hat{a}_i$ and $\hat{a}_i^\dagger$  are the harmonic oscillator annihilation and creation operators respectively and $\xi= \sqrt{2E_{\rm C \Sigma}/E_{\rm J}}$. Since the variable $\widetilde{\phi}_1$ is 2$\pi$ periodic, its conjugate operator $\hat{n}_1$ has discrete spectrum. On the other hand, the spectra of $\hat{n}_2$ and $\hat{n}_3$ are continuous.

\begin{figure*}[t!]
    \centering
    \subfloat[]{\includegraphics[width=0.33\textwidth]{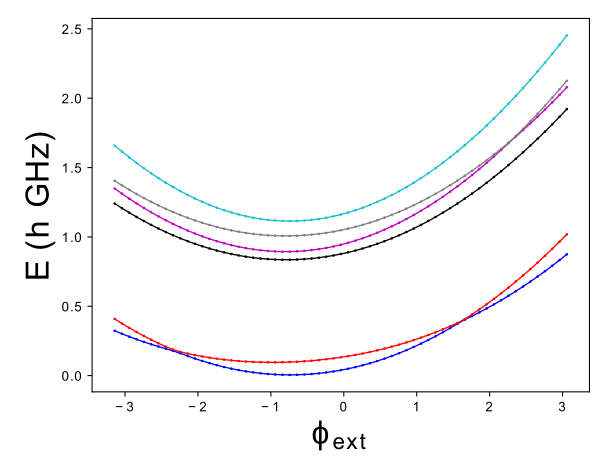}} 
    \subfloat[]{\includegraphics[width=0.33\textwidth]{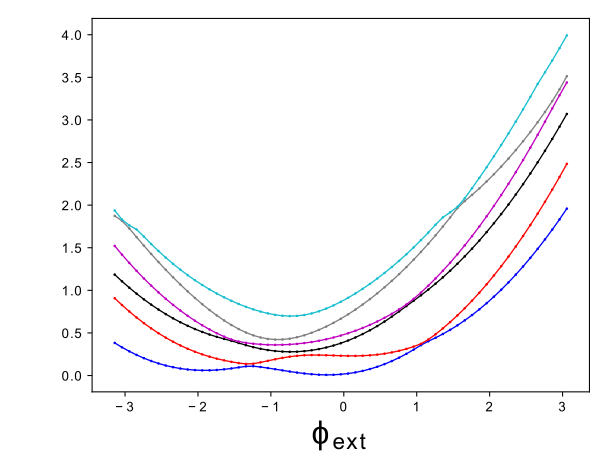}} 
    \subfloat[]{\includegraphics[width=0.33\textwidth]{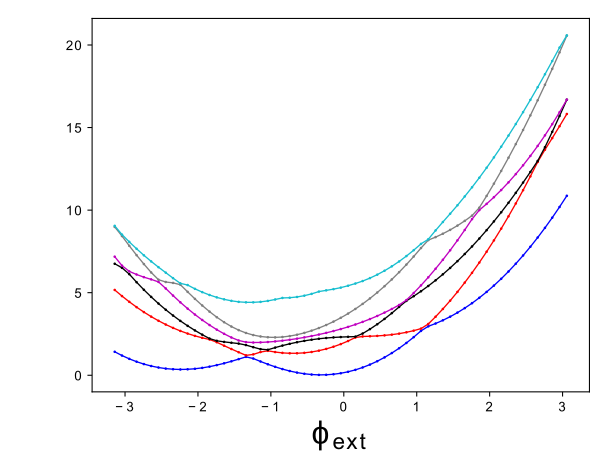}}
    \caption{The energy eigenspectrum for the Hamiltonian, Eq \eqref{eq22}, against the external flux $\phi_{\rm ext}$ for the three Sets of parameters (in units of h GHz): (a) $E_{\rm C} = 1.2$, $E_{\rm CJ} = 4$, $E_{J} = 6$, $E_L= 0.038$, (b) $E_{\rm C} = 0.15$, $E_{\rm CJ} = 10$, $E_{J} = 5$, $E_L= 0.13$, (c) $E_{\rm C} = 0.65$, $E_{\rm CJ} = 1.75$, $E_{J} = 10.8$, $E_L= 0.79$.} 
    \label{fig3}
\end{figure*}

For the potential term \eqref{eq13}, the cosine terms can be written in the operator form as
\begin{equation}\label{eq24}
    \cos \hat{\widetilde \phi}_i = \sum_{i= -N}^{N-1} \frac{1}{2} \big(\ket{i}\bra{i+1}+ \ket{i+1}\bra{i}\big),
\end{equation}
where $\ket{i}$ represent the eigenstates of the charge operator $\hat{n}_i$ and $N$ is the total number of charge states. The Hamiltonian \eqref{eq22} can thus be written in the charge basis $\{\ket{n_1, n_2,n_3}\}$ and its diagonalization gives the eigenvalues which are plotted in Fig. \ref{fig3} against the external flux $\phi_{\rm ext}$, for three different parameter sets where we have assumed $\phi_{\rm ext} = \phi_{\rm ext, i},~ i=1,2,3$.

In particular, for the parameter set (a) we get the qubit levels which are well separated from the higher energy levels which gives the system a high anharmonicity. Moreover, in this case the system exhibits avoided level crossings for $\phi_{\rm ext}= -2.3$ and  $\phi_{\rm ext}= 1.6$. From here on, we consider this parameter set (a) in Fig. \ref{fig3} for further study of this system. The avoided level crossings enable the non-adiabatic transitions between two states. The probability of these transitions between two energy states can be given by the Landau–Zener transition probability formula. But the points besides the avoided crossing are the no transition points and as desirable, these points are protected from the noise fluctuations.

\section{Flux Noise and qubit relaxation}\label{sec:noise}
\begin{figure*}[t!]
    \centering
    \subfloat[]{\label{fig4a} \includegraphics[width=0.45\textwidth]{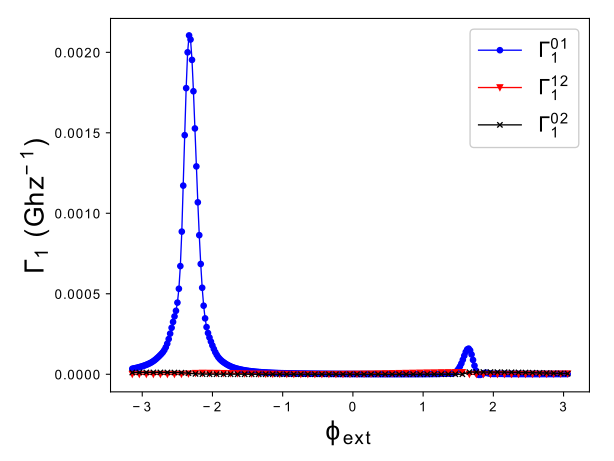}} 
    \subfloat[]{\label{fig4b} \includegraphics[width=0.45\textwidth]{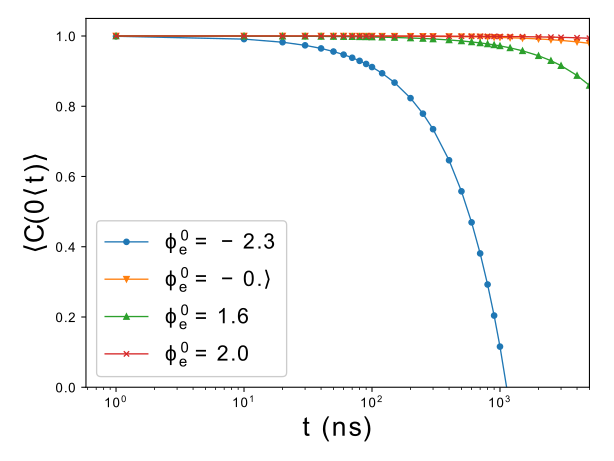}}
    \caption{\small{(a) Relaxation rate (Eq. \eqref{eq25}) for the parameter set (a) in Fig. \ref{fig3}} with time independent part of $\Phi_{\rm ext}$, $\phi_{\rm e}^0$, shows clearly that the transition rate is maximum between the lowest two levels at the two points of avoided level crossings. (b) The noise averaged qubit relaxation probability from the first excited state to ground state due to flux noise is shown here against time.} 
\end{figure*}

Let us now consider the response of the $0-\pi$ qubit in the presence of time-dependent external flux. In general, the perturbation will result in transitions between the qubit states. The transition rate between the states $\ket{m}$ and $\ket{n}$ of the Hamiltonian can be expressed as \cite{messiah}:
\begin{equation}\label{eq25}
    \Gamma_{mn} = \frac{1}{\hbar^2}\big|\braket{m|\sum_{i} \partial_{ {\phi}_{\rm ext, i}}\mathcal{H}(t)|n}\big|^2 S(\omega_{mn}),
\end{equation}
where $\omega_{mn}$ is the transition frequency between the qubit levels and $S(\omega_{mn})$ is a spectral function of flux noise at frequency $\omega_{mn}$, and for simplicity we consider flux noise to be Gaussian with short correlation time. In Fig. \ref{fig4a} we plot the relaxation rate for the lowest three energy levels of the unperturbed Hamiltonian w.r.t. the flux parameter $\phi_{\rm e}^0$. Whereas the transitions between the states $\ket{2}$ and $\ket{1}$, as well as between $\ket{2}$ and $\ket{0}$ are almost negligible, we see enhanced non-adiabatic transitions between $\ket{1}$ and $\ket{0}$ due to the presence of avoided level crossings.

If the qubit is prepared in the state $\ket{1}$, the flux noise will result in qubit relaxation to the ground state at a rate characterized by $\Gamma_{10}$. For the qubit relaxation probability due to flux noise, one obtains the averaged result \cite{you2019circuit}
\begin{equation}\label{eq26}
    \langle C(0,t) \rangle = \langle | \braket{\psi(0)| \psi(t)}|^2 \rangle \approx 1- \Gamma_{10} t.
\end{equation}
In Fig. \ref{fig4b} we plot the relaxation probability \eqref{eq26} for different values of the parameter $\phi_{\rm e}^0$. The plot gives us the flux values where the relaxation time scale $T_1$ of the qubit is maximum, which is desirable from an experimental perspective. Clearly, the flux must be tuned  away from points of the avoided level crossings to obtain maximum relaxation time. In particular for flux values of $\phi_{\rm e}^{0}= -0.5,2.0$ the relaxation time $T_1 \sim 10$ $\mu$s which can be seen if we extrapolate the curve of Fig. \ref{fig4b} to larger $t$ values.

\section{Survival and transition probabilities}\label{sec:survival}

\begin{figure}[b!]
    \begin{center}
    \includegraphics[width=0.8\textwidth]{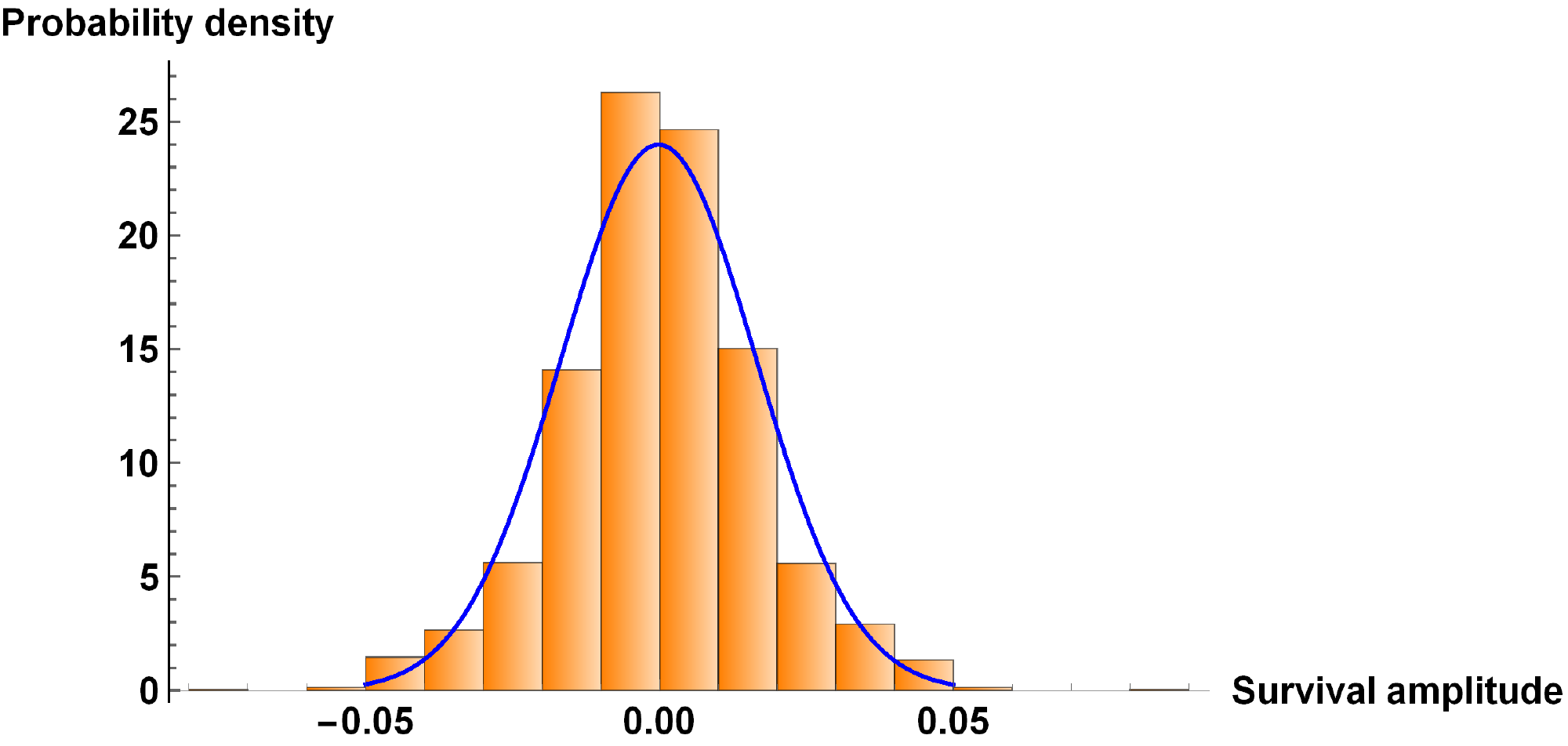}
    \caption{An initial state is taken as a superposition of all the 1331 states. The survival probability decays in time, with its real and imaginary parts also decaying individually. After the initial decay, the quantum fluctuations in the survival amplitude contains the spectral information of the quantum system. The probability density of the real part of the amplitude is shown here. The histogram fits very well with a Gaussian distribution with mean ($\mu$)=$-0.000158577$ and standard deviation ($\sigma$)=$0.016629$, demonstrating the complex nature of the underlying energy spectrum. It is not surprising as the underlying classical system is nonlinear and non-integrable. However, this quantitative measure of the complexity is important and instructive.}
    \label{samp}
    \end{center}
\end{figure}

Here we consider the evolution of an initial state under the time-dependent Hamiltonian in generality. For a quantum system prepared in a state, upon evolution in time, the survival amplitude is given by
\begin{equation}\label{eq:survive}
S(t) = \langle \psi |U(t)| \psi \rangle .     
\end{equation}
To put this quantity in a slightly larger perspective, let us note that it is related to a two-time correlation function at $\beta = 1/k_BT = 0$ in statistical mechanics of a system approaching equilibrium while responding linearly to an external perturbation. Due to the fact that the qubit system is nonlinear and classically non-integrable, the survival amplitude decays as an almost periodic function. If we consider only a few energy levels by truncating the Hilbert space, the amplitude will be a quasiperiodic function. However, we consider the entire spectrum (1331 energy levels and states for charge state, 11), the sum in \eqref{eq:survive} is more complex but not a random function, hence the term, almost periodic function \cite{jessen1945mean}.

This fine distinction can be brought out in the probability distribution function of the real or imaginary parts of the survival amplitude. Due to the results found by Pearson, Rayleigh, and Kluyver, for a set of energy levels which are statistically independent, the probability density can be shown to be Gaussian \cite{watson1995treatise}. For the system considered here, the distribution function of the real parts of the survival amplitude at different times is shown in Fig. \ref{samp}. It fits very well to a Gaussian distribution.   

The survival probability, $\mathcal{S}(t)$:
\begin{alignat}{1}
{\mathcal S}(t) &= |S(t) |^2 \nonumber \\
&= |\langle \psi (0)|e^{\frac{\iota}{\hbar}\int_{0}^{t} H(t^\prime)dt^\prime} | \psi (0)\rangle |^2.    
\end{alignat}
An initial state $\psi_i(0)$ of the system, under the influence of a time-dependent Hamiltonian spreads to $\psi_j (t)$. The transition probability is defined as
\begin{alignat}{1}
\mathcal{P}_r(t) &= |\langle \psi_j (t) | \psi_i(0)\rangle |^2 \nonumber \\   
&= |\langle \psi_j (0)|e^{\frac{\iota}{\hbar}\int_{0}^{t} H(t^\prime)dt^\prime} | \psi_i (0)\rangle |^2.    
\end{alignat}

\begin{figure}[b!]
\centering
    {\includegraphics[width=0.5\textwidth]{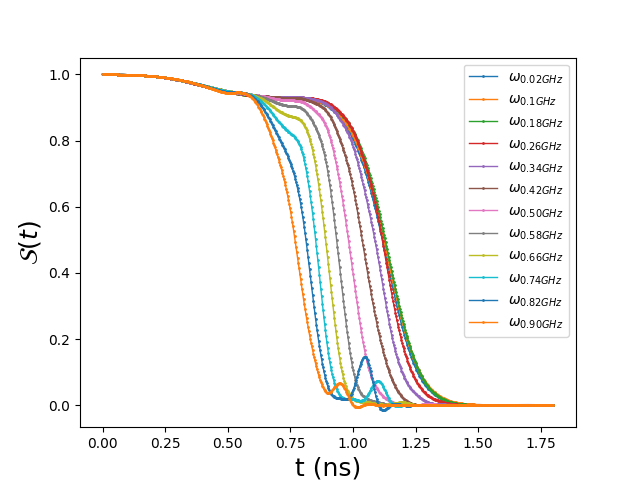}}
      \caption{With an external flux of a cosine form \eqref{flux}, and the junction parameters from the Set 1 from Table \ref{tab1}, we see here the the  survival probability of an initial ground state,  $|\psi\rangle_{in}=|\psi_0\rangle$ as a function of time. At a fixed moderate strength, $a_i=0.5$, for different driving frequencies $\omega_i$,  the state survives longest for the smallest frequency. As the frequency increases, the faster oscillation perturbs the system away from its initial state. }
    \label{sp.png}
\end{figure}

The total Hamiltonian $H(\vec{\widetilde\Phi},\vec{q},t)$ (Eq. \eqref{eq16}) for this system is time-dependent and the evolution is found using the Magnus expansion \cite{tannor2007introduction}. We consider the qubit to be coupled to a cavity which is driven by an external flux which has a simple time-dependent  form:
\begin{equation}\label{flux}
    \delta\Phi_{\rm e, i}(t)=a_i \Phi_0\cos (\omega_i t),
\end{equation}
with the corresponding phase variable, $\delta\phi_{\rm e, i}(t)=a_i \cos (\omega_i t)$, where $a_i$ is a constant, signifying the strength of the driving flux. Preparing the system in an initial ground state, survival probability gives a measure of the time up to which the system can remain in this state. The dynamics of this driven, classically nonlinear system sensitively depends on the parameters, strength and driving frequency. We have studied for a wide range of parameter values and noted that for certain strength, the system stays in the initial state for a relatively longer time. Recall that this is now a statement when we are driving the $0-\pi$ qubit. 

For concreteness, let us choose an  initial state to be one of the lowest eigenstates of the unperturbed system: $|\psi\rangle_{in}=|\psi_0\rangle$ at $E_1-E_0=0.0919945$ h-GHz, or, $|\psi_1\rangle$ at $E_2-E_1=0.74615$ h-GHz, the corresponding frequencies between first two lowest states, and, second two lowest states respectively, for different values of $a_i$ ($=0.1, 0.5$).  

Choosing the junction parameters from the Set 1 (Table \ref{tab1}), we show the survival probability of the ground state in Fig. \ref{sp.png} for frequencies ranging from 0.02 GHz to 0.9 GHz. As the frequency increases, the external flux samples the system a larger number of times, thus driving it out of its initial state earlier. Due to an explicit time-dependence, of course, the system stays in the ground state for a shorter period, 1 ns. However, we may prolong this time by employing the ideas from weak measurement theory and quantum Zeno effect \cite{kumar2020engineering}.

\begin{figure*}[t!]
    \centering
    \subfloat[]{\label{fig7a} \includegraphics[width=0.52\textwidth]{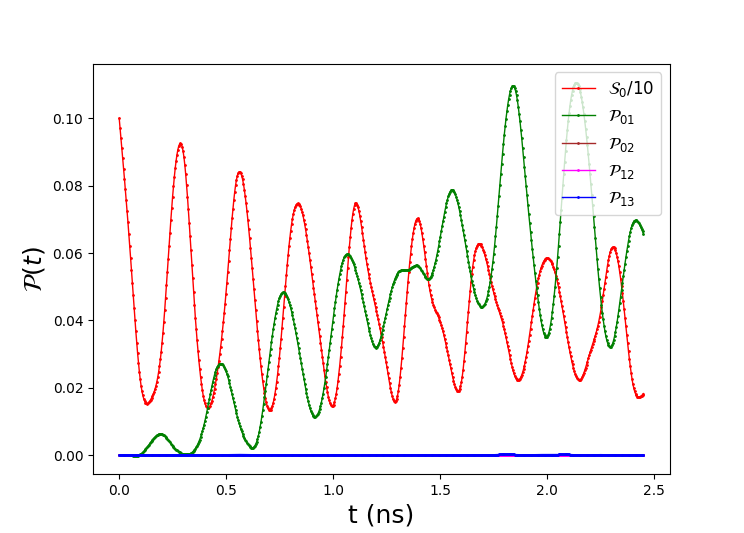}}
    \subfloat[]{\label{fig7b} \includegraphics[width=0.49\textwidth]{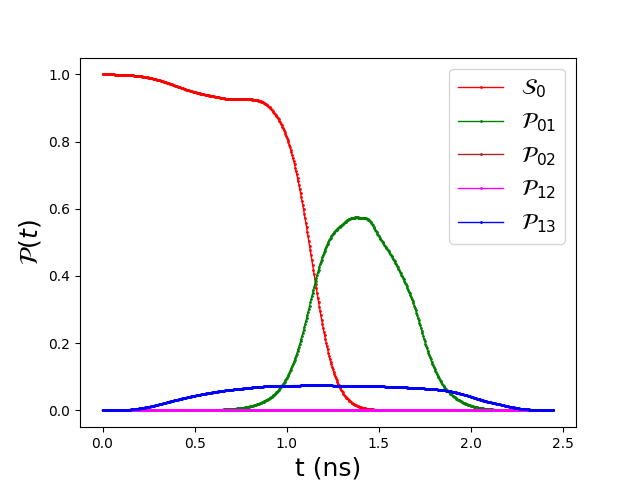}}
      \caption{For the same junction parameters as in Fig. \ref{sp.png}, and $\omega_i = 0.092$ GHz $\simeq  ~\omega_{01}$, (a) the  survival probability at a weaker strength ($a_i=0.1$) of  an initial ground state is shown along with the transition probabilities ${\mathcal P}_{01}$ (from state $|0\rangle\to|1\rangle$), ${\mathcal P}_{02}$ (from state $|0\rangle\to|2\rangle$), ${\mathcal P}_{12}$ (from state $|1\rangle\to|2\rangle$) and ${\mathcal P}_{13}$ (from state $|1\rangle\to|3\rangle$). This shows that there are oscillations between the first two states, establishing that qubit remains protected in the presence of drive. (b) At a larger driving strength $a_i=0.5$, the system stays in the initial ground state for over a nanosecond before the oscillations between the first two states begin to dominate.}
    \label{sptp.png}
\end{figure*}

For weaker strength, in Fig. \ref{fig7a}, we observe that the survival probability and transition probability ${\mathcal P}_{01}$ are oscillating out of phase. This shows that there are oscillations between the lowest two states. Thus, the protected qubit performs coherent oscillations, with no other levels participating. In Fig. \ref{fig7b}, for the same strength as in Fig. \ref{sp.png}, we see a similar phenomenon. However, there also appear weak oscillations among other levels. 

For completeness, we would like to briefly summarize the results obtained for the Sets 2 and 3 (Table \ref{tab1}). 
The coherent oscillations last for a very short time for the parameters in Set 2, and the best survival time seems to be only a fraction of a nanosecond. For the Set 3, there are no oscillations any more, due obviously to the number of avoided crossings and closely lying states, leading to a destructive interference.   

\section{Concluding remarks}

We have presented a Hamiltonian description for a driven $0-\pi$ qubit which allows tunability along with protection, following the important works in \cite{brooks2013protected,you2019circuit}. Formulating it as time-dependent perturbation over a time-independent system, the evolution of energy levels, transition probabilities and relaxation rates are calculated using the parameters taken from different theoretical and experimental realizations of the $0-\pi$ qubit. The presence or absence of avoided level crossings in the energy spectrum indicates the desirable flux values where the relaxation time is large. In particular, for suitably tuned flux values, relaxation time of $\sim 10$ $\mu$s can be achieved.

The solutions of the time-dependent Schr\"{o}dinger equation when the qubit is driven shows that an appropriate choice of strength and driving frequency can help us control it. There has been a lot of work in the recent times where quantum Zeno effect and ideas from weak measurement theory have been employed to ``control" the quantum jumps \cite{minev2019catch,kumar2020engineering}. We believe that our results, based on a systematic and rigorous treatment, have laid the foundation for setting up control and tunability of multi-loop quantum circuits, $0-\pi$ being an archetypal example. 

The treatment of the $0-\pi$ qubit in terms of a multi-loop system also admits immediate generalization as already brought out by \cite{you2019circuit}. Employing the ideas from linear response theory, electrical conductivity of this system has been expressed in terms of time correlation functions of current-loops, thus presenting the Ohm's law \cite{preprint2021}. 

We would like to remark that it has been shown recently that the protection of a qubit or even a more complicated system may be ensured by operating it close to an elliptic point, in the vicinity of a nonlinear resonance \cite{saini2020protection}, which is almost always present in the classical phase space. This is also valid for the $0-\pi$ qubit also as the Hamiltonian is that of coupled nonlinear oscillators for which the results of Kolmogorov-Arnold-Moser theorem hold good \cite{lichtenberg2013regular}. However, an explicit time-dependence increases the number of frequencies available to satisfy the condition of a nonlinear resonance. It would be worthwhile to study the classical dynamics of this system in detail and determine the values of parameters corresponding to protection.  

\section*{Acknowledgments}

The authors are thankful to the Referee for helpful comments and suggestions.

\appendix
\section{Calculation of the Lagrangian} \label{appex-1}

The irrotational degrees of freedom are obtained by demanding that all terms in the Lagrangian $\propto\dot\Phi_{\rm ext}$ vanish. This leads to the condition \cite{you2019circuit}: 
\begin{equation}\label{eqa1}
    \textbf{R} \textbf{C}^{-1}\textbf{M}^{\rm T} =0.
\end{equation}
Consistent with this condition, we find a particular solutions for the matrix $\textbf{M}$:
\begin{align}\label{eqa2-1}
\overline{M}_{11}&=C_{\rm J}(C_{\rm L}+C_{\rm C})/(C_{\rm L}C_{\rm J}+C_{\rm J}C_{\rm C} +C_{\rm C}C_{\rm L})\nonumber\\
\overline{M}_{12}&=C_{\rm J}(C_{\rm L}+C_{\rm C})/(C_{\rm L}C_{\rm J}+C_{\rm J}C_{\rm C} +C_{\rm C}C_{\rm L})\nonumber\\
\overline{M}_{13}&=C_{\rm L}(C_{\rm J}+C_{\rm C})/(C_{\rm L}C_{\rm J}+C_{\rm J}C_{\rm C} +C_{\rm C}C_{\rm L})\nonumber\\
\overline{M}_{14}&=C_{\rm L}(C_{\rm J}+C_{\rm C})/(C_{\rm L}C_{\rm J}+C_{\rm J}C_{\rm C} +C_{\rm C}C_{\rm L})\nonumber\\
\overline{M}_{15}&=C_{\rm J}(C_{\rm J}-C_{\rm L})/(C_{\rm L}C_{\rm J}+C_{\rm J}C_{\rm C} +C_{\rm C}C_{\rm L})\nonumber\\
\overline{M}_{16}&=C_{\rm C}(C_{\rm L}+C_{\rm J}+2C_{\rm C})/(C_{\rm L}C_{\rm J}+C_{\rm J}C_{\rm C} +C_{\rm C}C_{\rm L})\nonumber\\
\end{align}
\begin{alignat}{1}\label{eqa2-2}
\overline{M}_{21}&=-C_{\rm J}(C_{\rm L}+C_{\rm C})/(C_{\rm L}C_{\rm J}+C_{\rm J}C_{\rm C} +C_{\rm C}C_{\rm L})\nonumber\\
\overline{M}_{22}&=C_{\rm J}(C_{\rm L}+C_{\rm C})/(C_{\rm L}C_{\rm J}+C_{\rm J}C_{\rm C} +C_{\rm C}C_{\rm L})\nonumber\\
\overline{M}_{23}&=-C_{\rm L}(C_{\rm J}+C_{\rm C})/(C_{\rm L}C_{\rm J}+C_{\rm J}C_{\rm C} +C_{\rm C}C_{\rm L})\nonumber\\
\overline{M}_{24}&=-C_{\rm L}(C_{\rm J}+2C_{\rm L}+3C_{\rm C})/(C_{\rm L}C_{\rm J}+C_{\rm J}C_{\rm C} +C_{\rm C}C_{\rm L})\nonumber\\
\overline{M}_{25}&=-C_{\rm C}(C_{\rm J}+C_{\rm L}+2C_{\rm C})/(C_{\rm L}C_{\rm J}+C_{\rm J}C_{\rm C} +C_{\rm C}C_{\rm L})\nonumber\\
\overline{M}_{26}&=-C_{\rm C}(C_{\rm J}+C_{\rm L}+2C_{\rm C})/(C_{\rm L}C_{\rm J}+C_{\rm J}C_{\rm C} +C_{\rm C}C_{\rm L})\nonumber\\
\end{alignat}
\begin{alignat}{1}\label{eqa2-3}
\overline{M}_{31}&=C_{\rm J}(C_{\rm C}-C_{\rm L})/(C_{\rm L}C_{\rm J}+C_{\rm J}C_{\rm C} +C_{\rm C}C_{\rm L})\nonumber\\
\overline{M}_{32}&=-C_{\rm J}(C_{\rm L}+C_{\rm C})/(C_{\rm L}C_{\rm J}+C_{\rm J}C_{\rm C} +C_{\rm C}C_{\rm L})\nonumber\\
\overline{M}_{33}&=-C_{\rm L}(C_{\rm J}+C_{\rm C})/(C_{\rm L}C_{\rm J}+C_{\rm J}C_{\rm C} +C_{\rm C}C_{\rm L})\nonumber\\
\overline{M}_{34}&=-C_{\rm L}(C_{\rm J}-C_{\rm C})/(C_{\rm L}C_{\rm J}+C_{\rm J}C_{\rm C} +C_{\rm C}C_{\rm L})\nonumber\\
\overline{M}_{35}&=C_{\rm C}(C_{\rm L}-C_{\rm J})/(C_{\rm L}C_{\rm J}+C_{\rm J}C_{\rm C} +C_{\rm C}C_{\rm L})\nonumber\\
\overline{M}_{36}&=-C_{\rm C}(C_{\rm J}+C_{\rm L})/(C_{\rm L}C_{\rm J}+C_{\rm J}C_{\rm C} +C_{\rm C}C_{\rm L})
\end{alignat}
Since we have assumed an auxiliary parallel capacitor across the inductor to obtain the irrotational degree of freedom, we can subsequently put $C_{\rm L}=0$. Thus the  matrix $M$ simplifies considerably:
\begin{align}\label{eqa3}
\overline{\textbf{M}}=\begin{pmatrix}1 & 1 & 0 & 0 & 1 & 1+2 \frac{C_{\rm C}}{C_{\rm J}} \\
-1 & 1 & 0 & 0 & -(1+2\frac{C_{\rm C}}{C_{\rm J}}) & 1+2 \frac{C_{\rm C}}{C_{\rm J}} \\
1 & -1 & 0 & 0 & -1 & -1 \\
1 & 0 & 1 & 0 & 0 & -1\\
0 & -1 & 0 & -1 & 0 & 1 \\
0 & 1 & -1 & 0 & 1 & 0 \\
\end{pmatrix}.
\end{align}
The set of all possible solutions of Eq. \eqref{eqa1} can be expressed as 
\begin{equation}\label{eqa4}
    \textbf{M} = \textbf{A} \overline{\textbf{M}},
\end{equation}
where $\textbf{A}$ is an arbitrary non singular $(N-F) \times (N-F)$ i.e. $3 \times 3$ matrix. 
To facilitate  numerical computation we chose the matrix $\textbf{A}$ such that the upper block diagonal of the matrix $\textbf{M}_{+}^{-1}$ has integer elements. To this end, we define the matrix $\textbf{A}$ as:
\begin{equation}\label{eqa5}
 \textbf{A}  = \begin{pmatrix}
  C_{\rm J}/2C_{\Sigma} & 0  & 0 \\
    0 & C_{\rm J}/4C_{\Sigma}  & C_{\rm J}/4C_{\Sigma}  \\
   0 & 0  & 1/2  \\
   \end{pmatrix},
\end{equation}
where $C_{\Sigma} = (C_{\rm C}+C_{\rm J})$.  Thus the matrix  $\textbf{M}$ in Eq. \ref{eqa4} becomes:
\begin{alignat}{1}\label{eqa6}
\textbf{M} = \begin{pmatrix}
  \frac{C_{\rm J}}{2C_{\Sigma}} & \frac{C_{\rm J}}{2C_{\Sigma}}  & 0 & 0 & \frac{C_{\rm J}}{2C_{\Sigma}} & \frac{2C_{\rm C}+C_{\rm J}}{2C_{\Sigma}}\vspace{1mm}\\
    0 & 0  & 0 & 0 & \frac{-1}{2} & \frac{-1}{2} \vspace{1mm}\\
   \frac{1}{2} &\frac{-1}{2}  & 0 & 0 & \frac{-1}{2} & \frac{-1}{2} \\
   \end{pmatrix}
\end{alignat}$\\$
Consequently, the augmented matrix $\textbf{M}_{+}$  reads 
\begin{alignat}{1}\label{eqa7}
\textbf{M}_+ = \begin{pmatrix}
  \frac{C_{\rm J}}{2C_{\Sigma}} & \frac{C_{\rm J}}{2C_{\Sigma}}  & 0 & 0 & \frac{C_{\rm J}}{2C_{\Sigma}} & \frac{2C_{\rm C}+C_{\rm J}}{2C_{\Sigma}}\vspace{1mm}\\
    0 & 0  & 0 & 0 & \frac{-1}{2} & \frac{-1}{2} \vspace{1mm}\\
   \frac{1}{2} &\frac{-1}{2}  & 0 & 0 & \frac{-1}{2} & \frac{-1}{2} \vspace{1mm}\\
    1 & 0 & 1 & 0 & 0 & -1  \\
  0 & -1 & 0 & -1 & 0 & 1\\
  0 & 1 & -1 & 0 & 1 & 0\\
   \end{pmatrix}
\end{alignat}
We can now calculate $\textbf{C}_{\rm eff}$ using  Eq. \eqref{eq8}
\begin{alignat}{1} \label{eqa8}
 \textbf{C}_{\rm eff} & = \begin{pmatrix}
  2C_{\Sigma} &  2C_{\Sigma}  & 0 & 0 & 0 & 0\vspace{1mm}\\
    2C_{\Sigma} &  4C_{\Sigma} &  -2C_{\rm J} & 0 & 0 & 0 \vspace{1mm}\\
    0 & -2C_{\rm J} & 2C_{\rm J} & 0 & 0 & 0 \vspace{1mm}\\
    0 & 0 & 0 & \frac{C_{\rm C}C_{\rm J}}{2C_{\Sigma}} & 0 & \frac{C_{\rm C}C_{\rm J}}{2C_{\Sigma}} \vspace{1mm}\\
    0 & 0 & 0 & 0 & 0 & 0 \vspace{1mm}\\
    0 & 0 & 0 & \frac{C_{\rm C}C_{\rm J}}{2C_{\Sigma}} & 0 & \frac{C_{\rm C}C_{\rm J}}{2C_{\Sigma}}
 \end{pmatrix}.
\end{alignat}
Finally using the above matrices $\textbf{M}_+$ and $\textbf{C}_{\rm eff}$  we can easily construct the Lagrangian equations \eqref{eq11} and \eqref{eq13}.

\section{The kinetic energy part of the Hamiltonian} \label{appex-2}
The constants $A$, $B$, $\dots$, $F$ in the kinetic energy part of the Hamiltonian $\mathcal{H}_k$ in Section \ref{sec-3} are given by
\begin{alignat}{1}\label{eqb1}
 A&=2C_{\rm J}(16C_{\rm C}^3+39C_{\rm C}^2C_{\rm J}+29C_{\rm C}C_{\rm J}^2+7C_{\rm J}^3) \nonumber\\
 B&=2C_{\rm J} C_\Sigma^2(C_{\rm C}+7C_\Sigma)\nonumber\\
 C&=C_\Sigma^2(14C_{\rm C}^2+27C_{\rm C}C_{\rm J}+14C_{\rm J}^2)\nonumber\\
 D&=-4C_{\rm J}C_{\Sigma}^2(C_{\rm C}+7C_{\Sigma})\nonumber\\
 E&= 28C_{\rm J}C_{\Sigma}^3\nonumber\\
 F&=-2C_{\rm J}C_{\Sigma}(14C_{\rm C}^2+29C_{\rm C}C_{\rm J}+14C_{\rm J}^2)\nonumber\\
\end{alignat}



\begin{thebibliography}{99}
\bibitem{nielsen2002quantum} M. Nielsen and I. Chuang, Quantum computation and quantum information, (American Association of Physics Teachers, 2002).

\bibitem{kitaev2002classical}A. Y. Kitaev, A. Shen and M. N. Vyalyi, Classical and quantum computation, (American Mathematical Soc., 2002).

\bibitem{preskill2018quantum}J. Preskill, Quantum Computing in the NISQ era and beyond, {\em Quantum} \textbf{2} pp. 79 (2018).

\bibitem{kitaev2006protected}A. Kitaev, Protected qubit based on a superconducting current mirror, {\em ArXiv Preprint Cond-mat/0609441} (2006).

\bibitem{brooks2013protected}P. Brooks, A. Kitaev and J. Preskill, Protected gates for superconducting qubits, {\em Physical Review A} \textbf{87}, 052306 (2013).

\bibitem{groszkowski2018coherence}P. Groszkowski, A. Di Paolo, A. Grimsmo, A. Blais, D. Schuster, A. Houck and
J. Koch, Coherence properties of the 0-$\pi$ qubit, {\em New Journal Of Physics} \textbf{20}, 043053 (2018).

\bibitem{gyenis2021experimental}A. Gyenis, P. S. Mundada, A. Di Paolo, T. M. Hazard, X. You, D. I. Schuster, J. Koch,
A. Blais and A. A. Houck, Experimental Realization of a Protected Superconducting Circuit Derived from the 0-$\pi$ Qubit. {\em PRX Quantum} \textbf{2}, 010339 (2021)

\bibitem{saini2020protection}R. K. Saini, R. Sehgal and S. R. Jain, Protection of qubits by nonlinear resonances, {\em arXiv Preprint arXiv:2011.10329} (2020).

\bibitem{douccot2012physical}B. Dou\c{c}ot and L. Ioffe, Physical implementation of protected qubits, {\em Reports On Progress In Physics}, \textbf{75}, 072001 (2012).

\bibitem{bell2014protected}M. T. Bell, J. Paramanandam, L. B. Ioffe and M. E. Gershenson, Protected Josephson rhombus chains, {\em Physical Review Letters}, \textbf{112}, 167001 (2014).

\bibitem{gladchenko2009superconducting}S. Gladchenko, D. Olaya, E. Dupont-Ferrier, B. Dou\c{c}cot, L. B. Ioffe and M. E. Gershenson, Superconducting nanocircuits for topologically protected qubits, {\em Nature Physics}, \textbf{5}, 48-53 (2009).

\bibitem{smith2020superconducting}W. Smith, A. Kou, X. Xiao, U. Vool and M. Devoret, Superconducting circuit protected by two-Cooper-pair tunneling, {\em Npj Quantum Information}, \textbf{6}, 1-9 (2020).

\bibitem{kalashnikov2020bifluxon}K. Kalashnikov, W. T. Hsieh, W. Zhang, W. S. Lu, P. Kamenov, A. Di Paolo, A. Blais, M. E. Gershenson and M. Bell: Fluxon-parity-protected superconducting qubit, {\em PRX Quantum}, \textbf{1}, 010307 (2020).

\bibitem{you2019circuit}X. You, J. A. Sauls and J. Koch, Circuit quantization in the presence of time-dependent external flux, {\em Physical Review B}, \textbf{99}, 174512 (2019).

\bibitem{dempster2014understanding}J. M. Dempster, B. Fu, D. G. Ferguson, D. Schuster and J. Koch, Understanding degenerate ground states of a protected quantum circuit in the presence of disorder, {\em Physical Review B}, \textbf{90}, 094518 (2014).

\bibitem{shen2015theoretical}F. Shen, Theoretical analysis of a protected superconducting qubit, (University of Waterloo, 2015).

\bibitem{vool2017introduction}U. Vool and M. Devoret, Introduction to quantum electromagnetic circuits, {\em International Journal Of Circuit Theory And Applications}, \textbf{45}, 897-934 (2017).

\bibitem{messiah}A. Messiah, Quantum Mechanics, (North Holland Publishing Company, Amsterdam, 1962).

\bibitem{jessen1945mean}B. Jessen and H. Tornehave, Mean motions and zeros of almost periodic functions, {\em Acta Mathematica}, \textbf{77}, 137-279 (1945).

\bibitem{watson1995treatise}G. N. Watson, A treatise on the theory of Bessel functions, (Cambridge university press, 1995).

\bibitem{tannor2007introduction}D. J. Tannor, Introduction to quantum mechanics: a time-dependent perspective (2007).

\bibitem{kumar2020engineering}P. Kumar, K. Snizhko and Y. Gefen, Engineering two-qubit mixed states with weak measurements, {\em Physical Review Research}, \textbf{2}, 042014 (2020).

\bibitem{minev2019catch}Z. K. Minev, S. O. Mundhada, S. Shankar, P. Reinhold, R. Guti\`{e}errez-J\'{o}auregui, R. J. Schoelkopf, M. Mirrahimi, H. J. Carmichael and M. H. Devoret, To catch and reverse a quantum jump mid-flight, {\em Nature}, \textbf{570}, 200-204 (2019).

\bibitem{preprint2021}G. Rajpoot, K. Kumari, S. Joshi and S. R. Jain, Electrical conductivity for the 0-$\pi$ qubit, {\em Preprint}, (2021).

\bibitem{lichtenberg2013regular}A. J. Lichtenberg and M. A. Lieberman, Regular and stochastic motion, (Springer Science and Business Media, 2013).

\end{thebibliography}
\end{document}